\documentclass[12pt]{article}
\usepackage{graphics,color}
\begin{document}
\thispagestyle{empty}
%\pagestyle{empty}
%\vspace*{2.5cm}
\noindent\
\\
\\
\\
\begin{center}
\large \bf  Excited Weak Bosons and Dark Matter
\end{center}
\hfill
 \vspace*{1cm}
\noindent
\begin{center}
{\bf Harald Fritzsch}\\
Department f\"ur Physik\\ 
Ludwig-Maximilians-Universit\"at\\
M\"unchen, Germany \\
\end{center}

\begin{abstract}

The weak bosons are bound states of new constituents. The p-wave excitations are studied.
The state with the lowest mass is identified with the boson, which has been discovered at the LHC. Specific properties of the excited bosons are discussed, in particular their decays into weak bosons and photons. The stable fermion, consisting of three haplons, provides the dark matter in our universe.     

\end{abstract}

\newpage

In the Standard Model of the electroweak interactions the weak bosons are elementary gauge bosons. The masses of the weak bosons and of the leptons and quarks are due to the  spontaneous breaking of the electroweak symmetry, which is generated by a scalar boson. The masses of the weak bosons cannot be calculated, since they depend on the vacuum expectation value of the scalar field. \\

In quantum chromodynamics the masses of the hadrons are due to the field energy of the quarks and gluons inside the hadrons. The hadron masses can be calculated in terms of the QCD scale parameter $\Lambda_c$, if the quark masses are neglected.\\

Thus in the Standard Theory there are two different ways to generate the masses of the particles. The masses of the baryons and mesons are due to the field energy of the gluons and quarks. The masses of the weak bosons, the leptons and the quarks are generated by the spontaneous symmetry breaking. \\ 

Here we discuss the possibility, that the masses of the weak bosons are also generated dynamically. This is possible, if the weak bosons are not elementary gauge bosons, but bound states, analogous to the $\rho$-mesons in QCD.\\

In 2012 a scalar boson with a mass of 125 GeV has been discovered at the LHC (ref.(1,2). This boson might be the boson, which is responsible for the generation of the masses of the weak bosons. But if the weak bosons are composite particles, the new  boson might be an excitation of the $Z$-boson. \\

We assume that the weak bosons consist of a fermion and its antiparticle, which are denoted as "haplons" (see also ref.(3,4,5,6,7)). The haplons are confined, as the quarks inside the hadrons. The new confining gauge theory is denoted as quantum haplodynamics ($QHD$). The $QHD$ mass scale is given by a mass parameter $\Lambda_h$, which determines the size of the weak bosons. \\

The haplons are massless and interact with each other through the exchange of massless gauge bosons. The number of these gauge bosons depends on the gauge group, which is unknown. We assume, that it is the same as the gauge group of $QCD$: $SU(3)$. Thus the binding of the haplons is due to the exchange of eight massless gauge bosons, which are the analogues of the gluons in quantum chromodynamics.\\ 

Two types of haplons are needed as constituents of the weak bosons, denoted by $\alpha$ and $\beta$. We assume that the electric charges of the haplons are the same as the charges of the quarks: (+2/3) and (-1/3). The three weak bosons have the following internal structure:
\begin{eqnarray}
W^+ = (\overline{\beta} \alpha),~~ W^- = (\overline{\alpha} \beta),~~W^3 =\frac{1}{\sqrt{2}} \left( \overline{\alpha} \alpha - \overline{\beta} \beta \right).
\end{eqnarray}
The $QHD$ mass scale can be estimated, using the observed value of the decay constant of the weak boson. This constant is analogous to the decay constant of the $\rho$-meson, which is directly related to the $QCD$ mass scale. The decay constant of the weak boson is given by the mass of the weak boson and the observed value of the weak mixing angle:
\begin{eqnarray}
F_{\rm W}=\sin\theta_{\rm W}\cdot \frac{M_{\rm W}}{e}.
\end{eqnarray}
We find $F_{\rm W} \simeq 0.125 ~{\rm TeV}$. Thus the $QHD$ mass scale is about 0.2 TeV, about thousand times larger than the $QCD$ mass scale (see also ref. (7,8)). \\

The weak bosons consist of pairs of haplons, which are in an s-wave. The spins of the two 
haplons are aligned, as the spins of the quarks in a $\rho$-meson. The first excited states are those, in which the two haplons are in a p-wave. The weak isospin and the angular momentum of these states are described by $(I,J)$.\\ 

There are three $SU(2)$ singlets S(0), S(1), S(2) and three $SU(2)$ triplets T(0), T(1), T(2). These bosons are analogous to the mesons in strong interaction physics, in which the quarks are in a p-wave. The scalar meson $\sigma$, the vector meson $h_1(1170)$ and the tensor meson $f_2(1270)$ are the $QCD$-analogues of the singlet states $S(0)$, $S(1)$ and $S(2)$. The isospin triplet mesons, the scalar meson $a_0(980)$, the vector meson $b_1(1235)$ and the tensor meson $a_2(1320)$, correspond to the bosons $T(0)$, $T(1)$ and $T(2)$.\\

What is the mass spectrum of these excited weak bosons? The boson $S(0)$ is the particle, discovered at CERN (ref. (1,2)) - thus the mass of $S(0)$ is about 125 GeV. In analogy to QCD we expect that the masses of the other p-wave states are between 0.2 TeV and 0.5 TeV. Here are the expected masses of the bosons $S(1)$ and $S(2)$:
\begin{eqnarray}
S(1)&: 0.32~ TeV \pm 0.06~ TeV,\nonumber\\
S(2)&: 0.34~ TeV \pm 0.06~ TeV.
\end{eqnarray}
The uncertainties of these masses are also taken from the uncertainties of the mesons in QCD. But this might be wrong - the uncertainties could be larger than indicated above.\\

The masses of the $SU(2)$-triplet bosons $T$ are slightly larger than the masses of the $S$-bosons: 
\begin{eqnarray}
T(0)&: 0.25~ TeV \pm 0.05~ TeV,\nonumber\\
T(1)&: 0.33~ TeV \pm 0.05~ TeV,\nonumber\\
T(2)&: 0.36~ TeV \pm 0.06~ TeV.
\end{eqnarray}
The $S(0)$-boson will decay into two charged weak bosons, into two $Z$-bosons, into a photon and a $Z$-boson, into two photons, into a lepton and an anti-lepton or into a quark and an anti-quark. Since the mass of $S(0)$ is less than twice the mass of the charged weak boson, one of the weak bosons must be virtual. Details were discussed in ref. 9. The expected decay rates are similar to the decay rates, expected for a Higgs boson. Here are the corresponding branching ratios: 
\begin{eqnarray}
S(0)\rightarrow WW: 0.22 \; ,\nonumber \\
S(0)\rightarrow ZZ: 0.027 \; ,\nonumber \\
S(0)\rightarrow Z\gamma: 0.001 \; ,\nonumber \\
S(0)\rightarrow \gamma\gamma: 0.003 \; .
\end{eqnarray}
Here we have assumed, that the $S(0)$ does decay into a quark and an antiquark or a lepton and an antilepton with the same rate as the Higgs boson, but we do not expect that the decay rates are given by the masses of the leptons or quark, as they would be, if the new boson would be the Higgs boson.\\  

The boson $S(1)$ is a vector boson, thus it cannot decay into two photons. I estimated the branching ratios for the decays of the $S(1)$-boson, taking into account the available phase space and assuming, that the mass of the $S(1)$-boson is 0.38 TeV:
\begin{eqnarray}
S(1)\rightarrow WW: 0.087 \; ,\nonumber \\
~S(1)\rightarrow ZZ: 0.026 \; ,\nonumber \\
~S(1)\rightarrow Z\gamma: 0.018 \; ,\nonumber \\
~S(1)\rightarrow WWZ: 0.060 \; ,\nonumber \\ 
~S(1)\rightarrow WW\gamma: 0.020 \; ,\nonumber \\
~S(1)\rightarrow ZZZ: 0.017 \; ,\nonumber \\
~S(1)\rightarrow ZZ\gamma: 0.017 \; ,\nonumber \\
~S(1)\rightarrow Z\gamma\gamma: 0.006 \; .
\end{eqnarray}
Here we assumed, that the branching ratio for decay into leptons or quarks is 75 \%. If this branching ratio is less than 0.75, the other branching rations would be larger.\\

Interesting are the decays of the neutral boson T(0,0). We assume that the mass of T(0,0) is 0.3 TeV and that the branching ratio of the decay into leptons and quarks is again 75 \%. Here we give only the branching ratios for the decays into two particles: 
\begin{eqnarray}
T(0,0)\rightarrow WW: 0.080 \; ,\nonumber \\
~T(0,0)\rightarrow ZZ: 0.022 \; ,\nonumber \\
~T(0,0)\rightarrow Z\gamma: 0.016 \; ,\nonumber \\
~T(0,0)\rightarrow \gamma\gamma: 0.003 \; .
\end{eqnarray}
The branching ratio for the decay of T(0,0) into two photons is comparable to the branching ratio for the decay of S(0) into two photons. Thus the neutral T(0) boson can be discovered at the Large Hadron Collider by observing two high energy photons. If the mass of the neutral T(0) boson is 0.3 TeV, the production cross section at the energy of 14 TeV should be about 20 \% of the cross section for producing the S(0) boson.\\ 

The charged bosons T(0,+) and T(0,-) can decay into a charged weak boson and a $Z$-boson or a photon. Here are the expected branching ratios, if we assume that these bosons decay also into fermions with a branching ratio of 75 \%:
\begin{eqnarray}
T(0,+)\rightarrow W(+)+\gamma: 0.04 \; ,\nonumber \\
~T(0,+)\rightarrow W(+)+Z: 0.10 \; .
\end{eqnarray}
We also expect that there exist fermions, composed of three haplons, in particular the two ground states:
\begin{eqnarray}
D^+=(\alpha\alpha\beta),\nonumber \\
D^0=(\alpha\beta\beta).
\end{eqnarray}

These fermions are the $QHD$-analogies of the proton and neutron in $QCD$. The masses of these particles are expected to be above 0.5 TeV and below 1 TeV. The haplons are massless, the mass of the charged D-fermion is slightly larger than the mass of the neutral D-fermion. The mass difference between the two D-fermions should be about 1 GeV. The charged fermion would decay into the neutral fermion, emitting a virtual weak boson, decaying into a muon or a positron and a neutrino. The lifetime of the charged fermion should be about $10^{-12}$ s. The neutral D-fermion has haplon number 3 and would be stable.\\

Shortly after the Big Bang the D-fermions and their antiparticles would be produced. Since there is a small asymmetry between the matter and the antimatter in the universe, due to the violation of the CP-symmetry, there would be more D-fermions than Anti-D-fermions. These would annihilate with the D-fermions, and finally the univserse would contain, besides protons and neutrons, a gas of stable neutral D-fermions, providing the dark matter in our universe.\\

The average density of dark matter in our galaxy is about $400~{\rm TeV}/{\rm m}^3$. If we assume, for example, that the mass of a D-fermion is 0.8 TeV, there should be $500~D-{\rm fermions}/{\rm m}^3$. The D-fermions should have a velocity of about 300 km/sec. Thus the momentum of a D-fermion is quite small, about 0.4 MeV. \\ 

A neutral D-fermion can emit a virtual Z-boson, which interacts with an atomic nucleus. The cross section for the reaction of a neutral D fermion with a nucleus should be about $10^{-44}$ cm$^2$.\\ 

In a collision of a neutral D-fermion with a nucleus one can observe in specific experiments the sudden change of the momentum of the nucleus. Thus the neutral D-fermions can be observed indirectly, e.g. in the experiemnts in the Gran Sasso Laboratory. The present experiments indicate that the mass of a neutral D-fermion should be larger 
than 0.4 TeV.\\

The D-fermions can be produced in pairs at the LHC. The production of a neutral D-fermion and its antiparticle could be observed, since a large amount of the energy would be missing. Thus far nothing has been observed in the experiments at the Large Hadron Collider.\\
 
If a charged D-fermion and its antiparticle are produced, it will decay inside the detector into a neutral D-fermion, emitting a charged electron or muon. Thus no charged particle can be observed, only the missing energy due to the emission of two neutral D-fermions.\\

\end{document}